\documentclass[apj]{emulateapj}
\usepackage{amssymb}
\usepackage{multirow}
\usepackage{graphicx}
\usepackage{subfigure}
\usepackage{amsmath}

\shorttitle{}
\shortauthors{Webster, Frebel \& Bland-Hawthorn}

\newcommand{\comments}[1]{}

\begin{document}

\title{Segue~1 -- A Compressed Star Formation History Before Reionization}

\author{David Webster}
\affil{Sydney Institute for Astronomy, School of Physics, University of Sydney, NSW 2006, Australia}
\email{d.webster@physics.usyd.edu.au}

\author{Anna Frebel}
\affil{Department of Physics and Kavli Institute for Astrophysics and Space Research, Massachusetts Institute of Technology, Cambridge, MA 02139, USA}

\author{Joss Bland-Hawthorn}
\affil{Sydney Institute for Astronomy, School of Physics, University of Sydney, NSW 2006, Australia}

\begin{abstract}

Segue~1 is the current best candidate for a ``first galaxy", a system which experienced only a single short burst of star formation and has since remained 
unchanged. Here we present possible star formation scenarios which can explain its unique metallicity distribution. While the 
majority of stars in all other ultra-faint dwarfs (UFDs) are within 0.5~dex of the mean [Fe/H] for the galaxy, 5 of the 7 stars in Segue~1 have a spread of 
$\Delta$[Fe/H]~$>0.8$~dex. We show that this distribution of metallicities canot be explained 
by a gradual build-up of stars, but instead requires clustered star formation. Chemical tagging allows the separate unresolved delta functions in abundance space to be associated with discrete events in space and time. This provides an opportunity to put the enrichment events into a time sequence and unravel the history of the system. We investigate two possible scenarios for the star formation history 
of Segue~1 using Fyris Alpha simulations of gas in a $10^7$~M$_\odot$ dark matter halo. The lack of stars with intermediate metallicities 
$-3<$~[Fe/H]~$<-2$ can 
be explained either by a pause in star formation caused by supernova 
feedback, or by the spread of metallicities resulting from one or two supernovae in a low-mass dark matter halo. Either possibility can 
reproduce the metallicity distribution function (MDF), as well as the other observed elemental abundances. 
The unusual MDF and the low luminosity of Segue~1 can be explained by it being a first galaxy that originated with $M_{\rm{vir}}\sim10^7$~M$_\odot$ at $z\sim10$.

\end{abstract}


\section{Introduction}

The formation and evolution of the first stars and galaxies can be studied through 
observations of nearby dwarf galaxies. Stars preserve the chemical signatures at the time of their formation, such that galaxies with an early 
truncation of their star formation history can be used to investigate the conditions of the high-redshift universe. These systems are called 
``first galaxies" or ``fossil galaxies" \citep{ricotti05,bovill09,bromm11,frebel12,frebel14, hawthorn15}, reflecting the idea that their stellar population 
has been frozen in time for $>10$~Gyr. 
The best candidates for such systems are the ultra-faint dwarfs (UFDs, $L<10^5$~L$_\odot$) discovered in the past decade. 
Most UFDs contain only old, metal-poor stars and appear 
to have simple star formation histories 
\citep{brown12,brown14}, making them excellent candidates for probing the chemical signatures of the first generations of stars. 
 
There is not yet a consensus on the definition of a ``first galaxy", but most authors exclude the $\sim10^6$~M$_\odot$ minihalos in which the first 
Population III stars formed and require a first galaxy to produce and retain a long-lived stellar system \citep{bromm11,hawthorn15}. 
\citet{frebel12} defined a first galaxy as one which had only a 
single short burst of star formation. Such a system would be enriched only by Population III stars and would therefore 
lack the signatures of Type Ia and AGB star enrichment events, resulting in enhanced alpha element abundances and low or zero neutron-capture element abundances. 

The best candidate to date for a first galaxy is Segue~1 \citep{frebel14}. Segue~1 was discovered by \citet{belokurov07}, and confirmed 
as a galaxy by \citet{geha09}, who showed that it was dark matter-dominated, as well as \citet{norris10}, who showed that it had a large spread in 
metallicity. The mean metallicity of Segue~1, [Fe/H]~$\approx-2.7$ made it the most metal-poor known galaxy, although the recently discovered 
Reticulum 2 has a similar mean metallicity \citep{koposov15b,simon15}. High resolution spectroscopy \citep{frebel14} showed that the 7 brightest red giants 
in Segue~1, with metallicities ranging from [Fe/H]~$=-3.8$ to $-1.4$, showed [$\alpha$/Fe]~$\approx0.5$ and suppressed [Sr/H] and [Ba/H], 
suggesting a maximum 
of one r-process or weak s-process event. This implies that no stars formed from gas enriched by AGB stars, suggesting that star formation in Segue~1 lasted less than the lifetime of 
7-8~M$_\odot$ stars, $\sim 30-50$~Myr.

The metallicity distribution function (MDF) of Segue~1 is also unusual. While all other known UFDs have metallicity distributions that peak near the mean 
metallicity and have only a tail of metal-poor or metal-rich stars, Segue~1 shows three stars with [Fe/H]~$\approx-3.6$, two with [Fe/H]~$\approx-2.4$ 
and two with [Fe/H]~$\approx-1.5$. Five of the seven stars are more metal-poor or metal-rich than 90\% of observed stars in the 6
UFDs observed by \citet{brown14}. In this work, we seek to explain the unusual MDF of Segue~1 by exploring possible scenarios for its star formation history. This can give us 
an insight into the conditions of star formation in the early universe, as well as suggesting new avenues for determining which systems are likely 
to be first galaxies.

In the local universe, star formation is observed only in clusters \citep{lada03}. There is some evidence to suggest that this should also 
be true for the high-redshift universe \citep{clark08, larsen12, karlsson12a, karlsson13}. Clusters have homogeneous abundances as long as they are not self-enriching, meaning that the timescale for its star formation must 
be less than the time of the first supernova. This is supported by observations of clusters, including open clusters \citep{desilva06,desilva07,feng14}.

The chemical homogeneity of clusters has led to the idea of chemical tagging \citep{hawthorn10a}, the search for 
stars with similar abundances in a range of elements, suggesting that they originated from the same cluster. Separate unresolved delta functions 
in chemical abundance space are associated with events which are discrete in space and time. This provides an opportunity to put these events into a 
time sequence and determine the detailed enrichment history of a system.

\citet{hawthorn10b} tailored chemical 
tagging to dwarf galaxies, and it was applied by \citet{karlsson12a} to Sextans, the best candidate for a cluster 
signature in a dwarf galaxy. They found that three stars with [Fe/H]~$=-2.7$ contained remarkably similar 
Mg, Ti, Cr and Ba abundances. If correct, this group of stars is more metal-poor than any other known star cluster in the Milky Way.

Recently, 17 ultra-faint satellites have been discovered in the Southern sky \citep{koposov15a,des15}. It is not yet confirmed whether they are all 
UFDs, as it is possible that some are globular clusters. The only systems to date with [Fe/H] data from spectroscopy are Reticulum 2 and 
Horologium 1, which are confirmed as UFDs from measurements of the velocity dispersion and the spread in [Fe/H] \citep{walker15,simon15,koposov15b}. Reticulum 2 is approximately tied with Segue~1 as the most metal-poor galaxy, with $\overline{\rm{[Fe/H]}}~=-2.7$. Reticulum 2 is one of the 6 objects described by \citet{koposov15a} 
as having sizes and luminosities similar to Segue~1. However, as we will see in Section~\ref{s:theomdfs}, the MDF of Reticulum 2 is more similar 
to the more luminous UFDs than to that of Segue~1. 

In Section~\ref{s:theomdfs} we discuss the MDFs produced under various assumptions for the star formation history of a UFD. We compare 
these to the MDFs of Segue~1, Reticulum 2, and the entire UFD population. In Section~\ref{s:scen}, we present possible star formation histories 
for Segue~1, guided by Fyris alpha hydrodynamical simulations of small (M$_{\rm{vir}}=10^7$~M$_\odot$) dark matter halos. We discuss our results in 
Section~\ref{s:disc} and summarise in Section~\ref{s:summary}.

\section{Theoretical MDF}
\label{s:theomdfs}


\subsection{Method}

In this section, we show MDFs resulting from various star formation histories and compare them to the MDF of Segue~1. Given that Segue~1 is 
believed to have formed stars for less than 50~Myr, it should have a relatively uncomplicated star formation history. 

\begin{figure*}
	\centering
	\includegraphics[width=.9\textwidth]{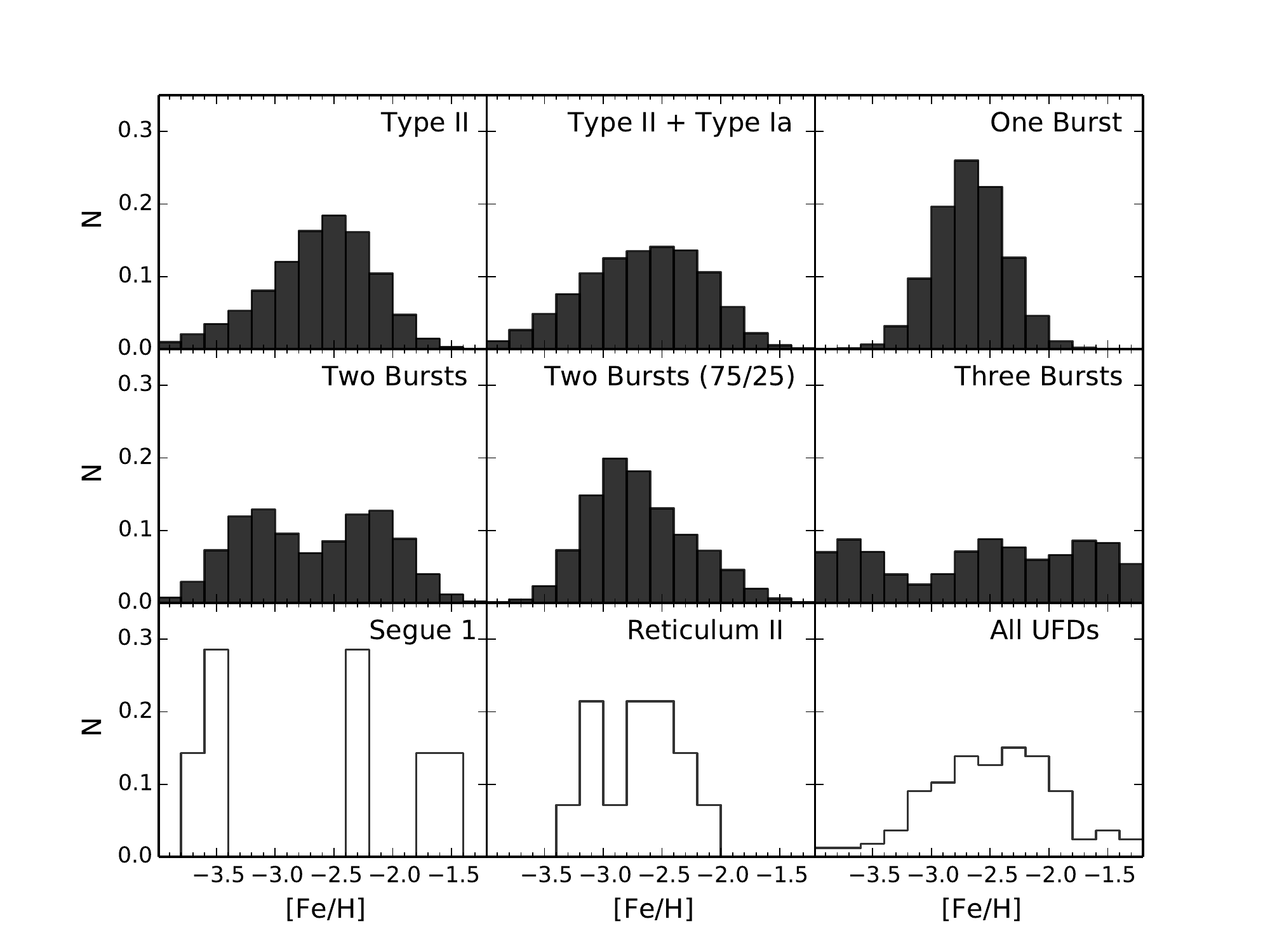}
	\caption{Theoretical normalised MDFs (filled histograms) resulting from a range of possible simple star formation histories, along with the normalised MDFs observed in UFDs (unfilled histograms). In the top six panels, the mean [Fe/H]~$\approx-2.7$, 
as in Segue~1 and Reticulum 2. The lower three panels show the observed MDFs of individual UFDs, as well as the entire observed UFD population \citep{kirby13,brown14,frebel14,simon15}.}
	\label{f:theoscen}
\end{figure*}

The first two panels of Figure~\ref{f:theoscen} correspond to models of continuous enrichment. In the first panel (labelled ``Type II") an equal amount of star formation occurs between each of 14 Type II supernovae. Between time $t$ and $t+\Delta t$, 100 stars form with [Fe/H] for a given star $s$ given by:

\begin{equation}
\begin{split}
[\text{Fe}/\text{H}](s) &= \text{log}(n_{\text{Fe}}(\text{s})/n_{\text{H}}) - \text{log}(n_{\text{Fe},\odot}/n_{\text{H},\odot}) \\
&=\text{log}(\frac{M_{\text{Fe}_i} + y_{\text{Fe}}N t(s)}{\mu_{\text{Fe}}M_\text{H}})- 4.55 + X
\end{split}
\end{equation}

where $M_{\text{H}} = 10^5~$M$_\odot$ and $M_{\text{Fe}_i}$, the initial mass of iron in the gas, is set such that 
[Fe/H] = $-3.5$ at $t=0$, $\mu_{\text{Fe}}$ is the atomic mass of iron, N is the number of supernova explosions per unit time, $y_{\text{Fe}}=0.08$~M$_\odot$ is the iron yield of 
each (Type II) supernova, $\text{log}(n_{\text{Fe},\odot}/n_{\text{H},\odot}) = -4.55$ \citep{asplund05} and $X\sim\mathcal{N}(0,0.3^2)$ is a random number selected from a Gaussian distribution with $\mu=0$, $\sigma=0.3$, representing the approximate dispersion in metallicities in the interstellar medium (ISM) \citep{feng14}. $N\Delta t$ is set to equal $1$, such that there is one supernova explosion between $t$ and $t+\Delta t$, the period during which 
100 stars form. The ``Type II+Ia" model in the second panel is the same, except that for a single value of $t$, taken to be half the total time, 
$y_{\text{Fe}} = 0.8$~M$_\odot$, corresponding to the yield of a Type Ia supernova. The supernova yields are from \citet{iwamoto99} and \citet{nomoto06}. The model was run 100 times and the average taken.

The neutral hydrogen mass $M_\text{H}=10^5$~M$_\odot$ assumed is of the order of the amount of gas in the inner 100~pc of an 
$M_{\rm{vir}}=10^{6.5-7}$~M$_\odot$ dark matter halo with a baryon fraction of 10\%. As shown in \citet{hawthorn15}, such a system can retain 
the majority of its gas and metals in the face of a single 
supernova explosion. The gas initial constant metallicity [Fe/H]~$=-3.5$, corresponds to the expected metallicity for such a system enriched by a first star \citep{greif07,frebel12}. At each timestep the Fe yield of a single Type II supernova is added to the 
total Fe of the gas, with the neutral hydrogen content assumed to remain constant, such that the metallicity of the gas gradually increases.

The second class of scenarios, shown in Panels 3-6 (``One Burst", ``Two Bursts", ``Two Bursts 75/25" and ``Three Bursts") of 
Figure~\ref{f:theoscen} are simply sums of Gaussian distributions. In all cases, the scatter of metallicities within a burst is given by 
the standard deviation of the Gaussians, set to be $0.3$~dex as for the continuous models.

This gives a total of 6 scenarios:
\begin{enumerate}
\item ``Type II": Continuous enrichment from Type II supernovae as in Equation 1.
\item ``Type II + Type Ia": Continuous enrichment from Type II and Type Ia supernovae as in Equation 1.
\item ``One Burst": A single burst of stars with mean [Fe/H]~$=-2.67$, represented by a single Gaussian.
\item ``Two Bursts": The sum of two equally-weighted Gaussian distributions with means [Fe/H]~$=-3.17$ and $-2.17$, representing the stars being evenly divided between the bursts.
\item ``Two Bursts 75/25": The sum of two Gaussians, one with mean [Fe/H]~$=-2.87$ and one with $-2.17$. The first Gaussian is given a relative weight of 
3 times the second, such that this is a model in which 75\% of the stars form in the lower metallicity burst.
\item ``Three Bursts": Three evenly-weighted Gaussians with mean [Fe/H]~$=-3.7$, $-2.5$ and $-1.6$, representative of three bursts with the stars evenly divided between the bursts.
\end{enumerate}

The first two scenarios correspond to closed box models, with no inflows or outflows, where the iron is mixed instantaneously with the neutral hydrogen gas. Star formation is terminated when $\overline{\rm{[Fe/H]}}$ in the stars formed is approximately $-2.7$, which is the mean [Fe/H] observed in Segue~1 and Reticulum 2. This number of supernovae is consistent with the 1500~M$_\odot$ of star formation implied for Segue~1 under a \citet{kroupa01} IMF \citep{frebel14}. Assuming a specific star formation rate of $\sim10^{-3}$~M$_\odot$yr$^{-1}$kpc$^{-2}$ \citep{bigiel08,webster14} as in the Carina dwarf, this will take $100-200$~Myr. 

The number of stars in a given [Fe/H] bin increases exponentially with increasing [Fe/H] until the mean metallicity at which star formation is switched off, because [Fe/H] is a logarithmic scale, and the Fe mass is assumed to increase linearly, with the hydrogen mass remaining constant. There are a few stars with higher metallicities, which result from the assumed dispersion of [Fe/H].

While Segue~1 shows no evidence of Type Ia enrichment, we include the second panel (``Type II + Type Ia") to show how Type Ia supernovae alter an MDF. The build-up of metallicities remains exponential for the low-metallicity stars which form first. However, a Type Ia 
supernova injects a large amount of Fe, resulting in a rapid increase in Fe which flattens the peak of the [Fe/H] distribution.

Scenarios 3-6 represent distinct bursts of star formation, with no self-enrichment within each burst and the gas considered to be 
well-mixed between each burst. They give an indication of the variety of qualitatively different 
MDFs which can be constructed using simple bursts. This is motivated by \citet{brown14}, who used isochrone fitting to determine the star formation history of 6 UFDs, finding that they were best fit by two bursts of star formation, although in 3 of the galaxies, $\geq95\%$ of the stars were in one of the bursts. Giving the bursts a non-zero duration did not improve the fit. Scenario 3 (``One Burst") has been proposed as a scenario for a first galaxy \citep{frebel12}, while the fourth scenario (``Two Bursts") is similar to the \citet{brown14} history for Ursa Major I. The fifth panel (``Two Bursts 75/25") shows two bursts with the first containing 75\% of 
the stars, similar to the \citet{brown14} star formation histories for Hercules and Leo IV.




The lower three panels of Figure \ref{f:theoscen} show data from observations of Segue~1 \citep{frebel14} and Reticulum 2 \citep{simon15}, as well as 
the combined data from the six UFDs observed by \citet{brown14} along with Segue 2 \citep{kirby13}, Segue~1 and Reticulum 2. While Reticulum 2 could 
be explained by a number of possible scenarios, Segue~1 is highly unusual. The closest match would be a three-burst scenario with smaller dispersion than 
shown in the ``Three Bursts" panel. The overall UFD population shares qualitative features with the continuous Type II + Ia model, but this could also 
be caused by it being a composite of a large number of bursts across many systems.

\subsection{Clustering in Segue~1} 

\begin{figure}
	\centering
	\includegraphics[width=.45\textwidth]{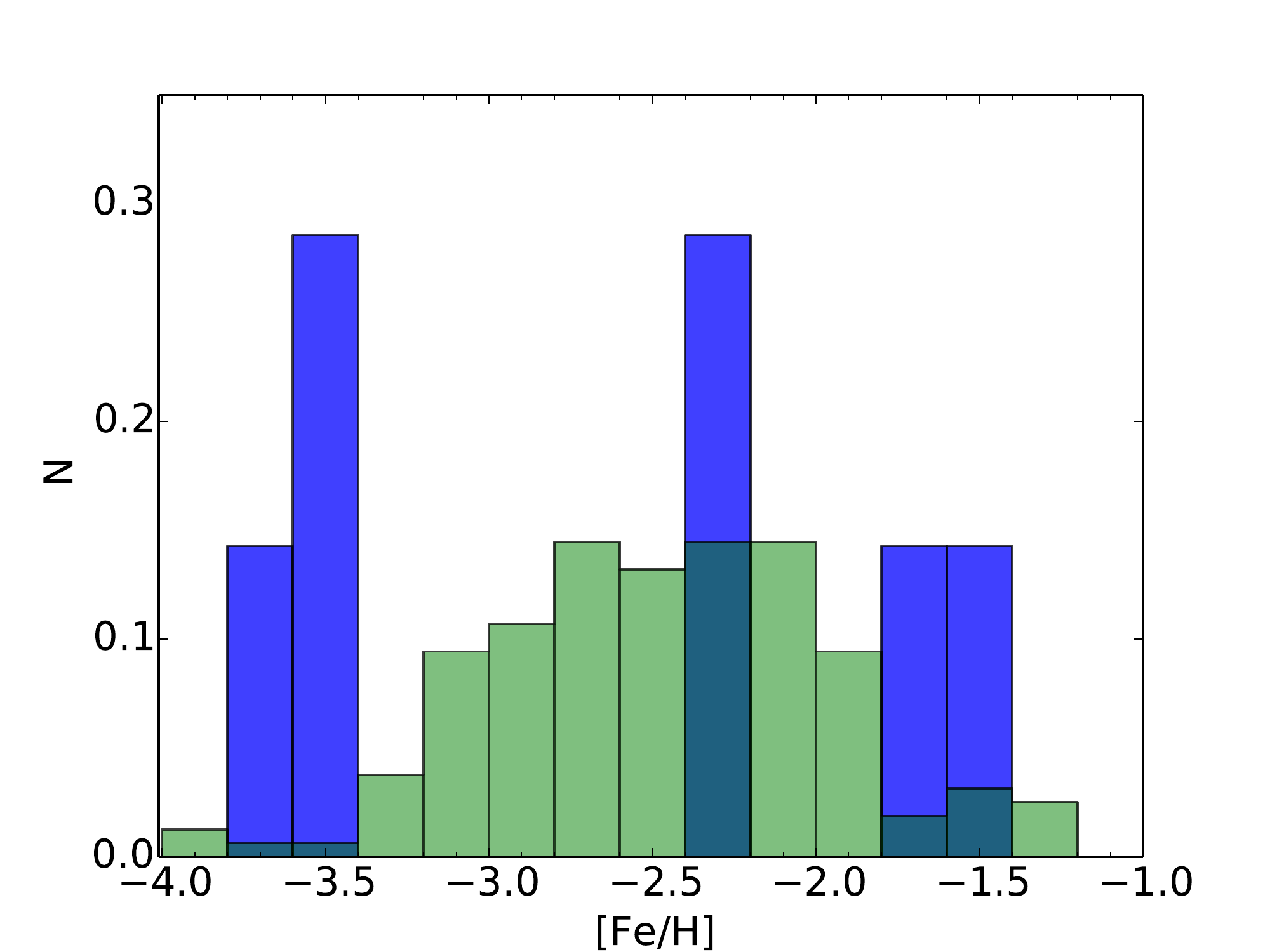}
	\caption{The normalised MDF of Segue~1 \citep{frebel14} (blue) and the combined normalised MDF of Bootes I, Coma Berenices, Canes Venatici II, Hercules, Leo IV and Ursa Major I \citep{brown14}, Reticulum 2 \citep{simon15} and Segue 2 \citep{kirby13} (green).}
	\label{f:segbrown}
\end{figure}

Figure~\ref{f:segbrown} shows how unusual Segue~1 is compared to the overall UFD population. For all UFDs except Segue~1, the majority of stars have 
metallicities within 0.5~dex of the mean. In Segue~1, 5 of the 7 stars have metallicities outside the range of 
$>90\%$ of the stars in UFDs. As well as the extreme metallicities, any scenario to explain the star formation history of Segue~1 needs to explain the gaps in metallicity between the three stars with [Fe/H]~$\approx -3.6$, 
the two with [Fe/H]~$\approx -2.4$ and the two with [Fe/H]~$\approx -1.5$. No such gap is seen anywhere else in MDFs of UFDs, with the exception of Canes 
Venatici II, which contains a single high-metallicity star. The favoured star formation history for Canes Venatici II from \citet{brown14} is two bursts, 
with the second burst separated from the first by 3.2~Gyr containing only 5\% of the stars, so this higher metallicity star is likely part of a small 
second burst. 

As a statistical test to compare a three cluster model to an unclustered model, we use the \citet{beale69} pseudo-$F$ statistic:

\begin{equation}
F^* = \frac{J_1^2-J_2^2}{J_2^2} \frac{(N-c_2)c_2^{-2/p}}{(N-1)-(N-c_2)c_2^{-2/p}}
\end{equation}
where $J_1^2 = \Sigma_i (x_i-\mu)^2$ and $J_2^2 = \Sigma_k\Sigma_{k_i} (x_{k_i} - \mu_k)^2$ correspond to the squared errors of the unclustered and clustered models, $\mu_k$ are the cluster centroids, $N$ is the number of data points, $c_2$ is the number of clusters and $p$ is the dimensionality of the data. The first term is a comparison of how well the data matches the models, while the second term includes only the number of clusters, number of 
observations and the dimensionality of the data, such that models with more parameters are penalised.

For a three-cluster model with centroids at ($-3.65, -2.36, -1.55$) compared to a model without clusters, $F^*= 274\approx F(2,4)_{p=0.0001}$, showing that the three-cluster model is a better fit to the data. The three-cluster model also outperforms the best two-cluster model, with 
$F^*=7\approx F(1,4)_{p=0.05}$.  



Gaps in the MDF can be produced by star formation in discrete bursts. Single-age bursts should produce groups of stars that form at about the same 
time and therefore have similar metallicities. However, even three bursts centered on the metallicity of the three groups in Segue~1 will not necessarily 
produce an MDF like in Segue~1. While there may be gaps in the MDF, typically one or more of the groups will contain either no stars or only one star, 
which does not conclusively show that the groups exist. Put differently, it is possible that there are systems with a similar star formation history to Segue~1 where we 
will not observe one or more of the groups and the system will therefore not be recognised as having grouped star formation.

There are good physical and observational reasons to expect that star formation should be clustered in time and/or place in low mass systems. 
In the local universe, stars are observed to form in clusters \citep{lada03}. If this is also the case at the very low gas masses and star formation rates of the UFDs, as suggested by \citet{clark08}, we would expect to observe groups of stars with the same chemical composition. In larger systems, 
this effect is washed out because eventually stars exist at all metallicities. However, in a ``first galaxy" such as Segue~1, there are no later generations, meaning that the signature of the first clusters are observed. As shown in \citet{hawthorn15} and \citet{webster14}, in a small enough 
dark matter halo, a single supernova can affect the entire system, with star formation only possible during sufficiently long breaks between supernovae. 
If there are a large number of supernovae during a short period of time, the gas may be blown out completely, permanently ending star formation in the 
system. 

The unusual MDF of Segue~1 is likely connected to its status as a first galaxy, with both features caused by its inability to retain gas for 
more than $\sim50$~Myr. This is plausible if Segue~1 had a mass $M_{\rm{vir}}\sim10^7$~M$_\odot$ when it formed stars at $z>10$, which would imply 
a current mass of $M_{\rm{vir}}=10^{8-9}$~M$_\odot$ \citep{webster15a}. For Segue~1, the observed half-light mass $M_{\rm{half}}=2\pm1\times10^5$~M$_\odot$
and half-light radius $r_{\rm{half}}\approx30$~pc \citep{collins14} give a best fit mass assuming an NFW profile $M_{\rm{vir}}\sim3\times10^{9}$~M$_\odot$, but with an uncertainty of more than an order of magnitude. 

Given that continuous enrichment is unlikely, Segue~1 must have formed its stars in separate bursts. However, these bursts can not have been similar to 
those determined by \citet{brown14}, which produced stars with a range of [Fe/H] and [$\alpha$/Fe] and likely had some self-enrichment \citep{webster15a}. 
For Segue~1, the three bursts each contain red giants which are consistent within the uncertainties with the stars within each cluster having 
the same [Fe/H]. 

\subsection{Carbon}

In the Milky Way halo and in UFDs, a large proportion of very metal-poor stars are enhanced in carbon relative to iron. This is seen in Segue~1, where 
4 of the 7 stars with measured abundances are carbon-enhanced metal-poor (CEMP) stars, with [C/Fe]~$>0.7$, although one of the four appears to have  received carbon from a binary companion \citep{frebel14}. This star is excluded from the discussion below. The scatter $\Delta$[C/Fe] is $>2$~dex, significantly larger than for heavier elements in Segue~1, suggesting that carbon enrichment was decoupled from the enrichment of heavier elements \citep[see also][]{frebel14,ritter15,sluder15}.  

\begin{figure}
     \centering
      \includegraphics[width=.45\textwidth]{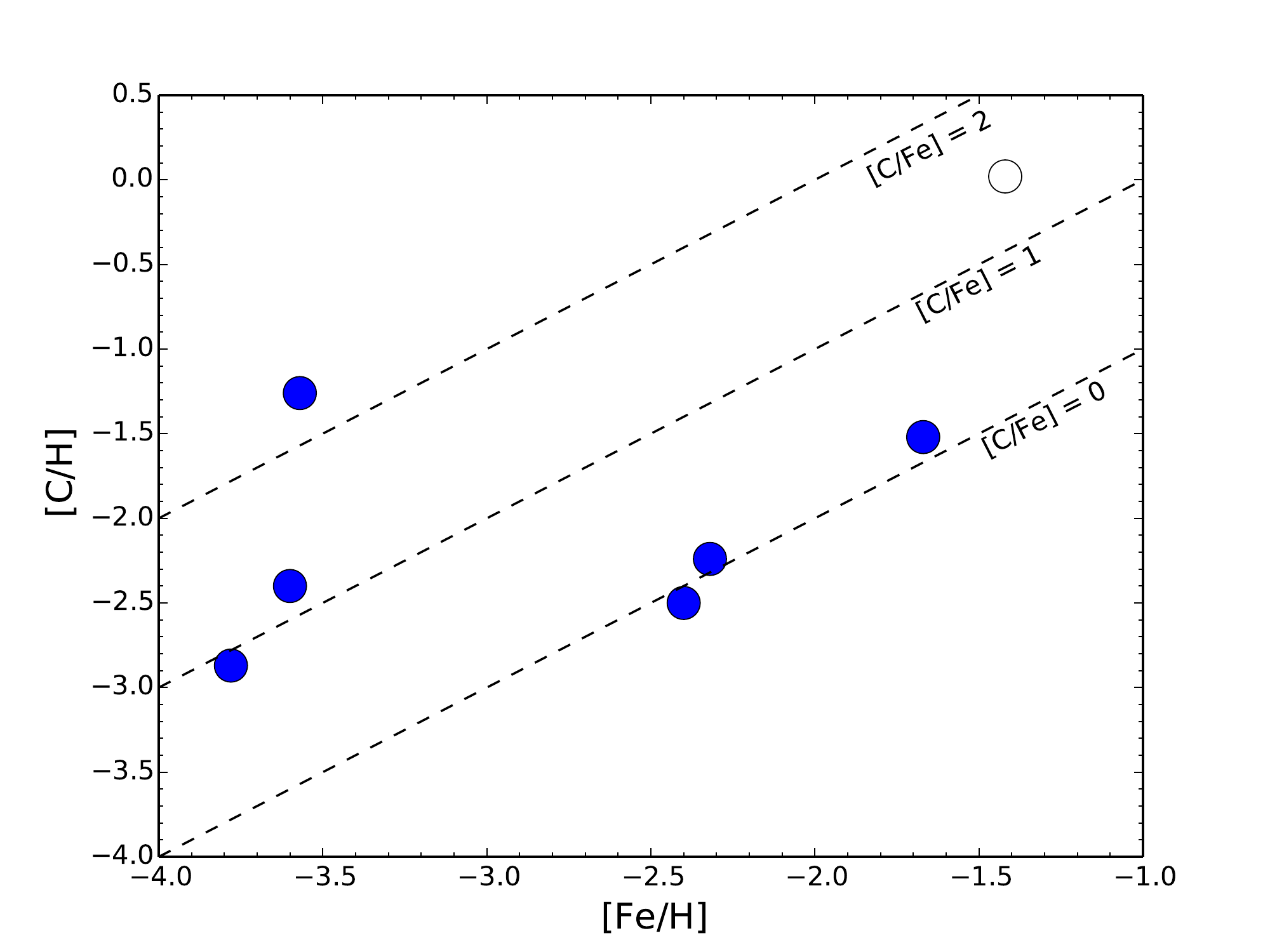}
      \caption{[C/H] vs [Fe/H] for Segue~1, with dashed lines denoting lines of constant [C/Fe]. The empty circle indicates the star which has been 
carbon-enhanced by its binary companion, while the filled in circles are stars which likely retain the abundance of the gas in which they formed. The data is from \citet{frebel14} and \citet{norris10}.}
     \label{f:cvfe}
\end{figure}

Figure~\ref{f:cvfe} shows the relationship between [C/H] and [Fe/H], along with lines representing constant [C/Fe]. There is no clear trend of [C/H] 
with [Fe/H]. This is suggestive that the stars with higher [Fe/H] in Segue~1 need not be a consequence of a large number of 
enrichment events, but are instead the result of inhomogeneous mixing. With few events, the inherent scatter that results from a single supernova becomes 
more important, as we will see in Section~\ref{s:scen}.

Figure~\ref{f:cvfe} also provides some additional evidence for clustered star formation in Segue~1 in that despite the high overall scatter in carbon 
abundances, the two stars with [Fe/H]~$\approx-2.4$ have similar [C/H], 
as do 2 of the 3 stars with [Fe/H]~$\approx-3.5$. The only outlier as classified by the Fe-clusters is the star with [C/H]~$= -1.25$ at [Fe/H]~$=-3.57$. 
This star is from a different observational 
sample to the others and is also an outlier in [Mg/Fe], [Al/Fe] and [Ni/Fe] \citep{norris10,frebel14}.

\citet{ritter15} explain stars with enhanced [C/Fe] as resulting from differences in the entropy of the inner and outer regions of the supernova ejecta. In 
their simulations, the gas which first returns to the center is deficient in the innermost 10\% of the ejecta, which 
contains iron peak and $\alpha$ elements. The outer ejecta cools first and does not rise to the same radii as the inner ejecta. The outer ejecta collapses 
to the center of the system, resulting in carbon-enhanced, iron-poor star formation. In Segue~1, there is low variance in [X/Fe] abundances of stars for 
$\alpha$ and Fe-peak elements, but [C/Fe] does vary significantly between clusters. This is consistent with the \citet{ritter15} single and clustered 
supernovae simulations.

\section{Scenarios for the star formation history of Segue~1}
\label{s:scen}

In the discussion above, we considered only the gaps in the MDF and were not concerned with the precise metallicities. We have not yet provided a 
solution as to how stars with metallicities as high 
as [Fe/H]~$\sim-1.5$ can form. No other UFD contains more than one observed star with [Fe/H] more than 1~dex higher than its mean [Fe/H]. 
This is despite the fact that unlike Segue~1, most of these systems likely had a longer star formation history, as they show evidence of enrichment 
from AGB stars and Type Ia supernovae.

\begin{figure*}
     \centering
      \includegraphics[width=.9\textwidth]{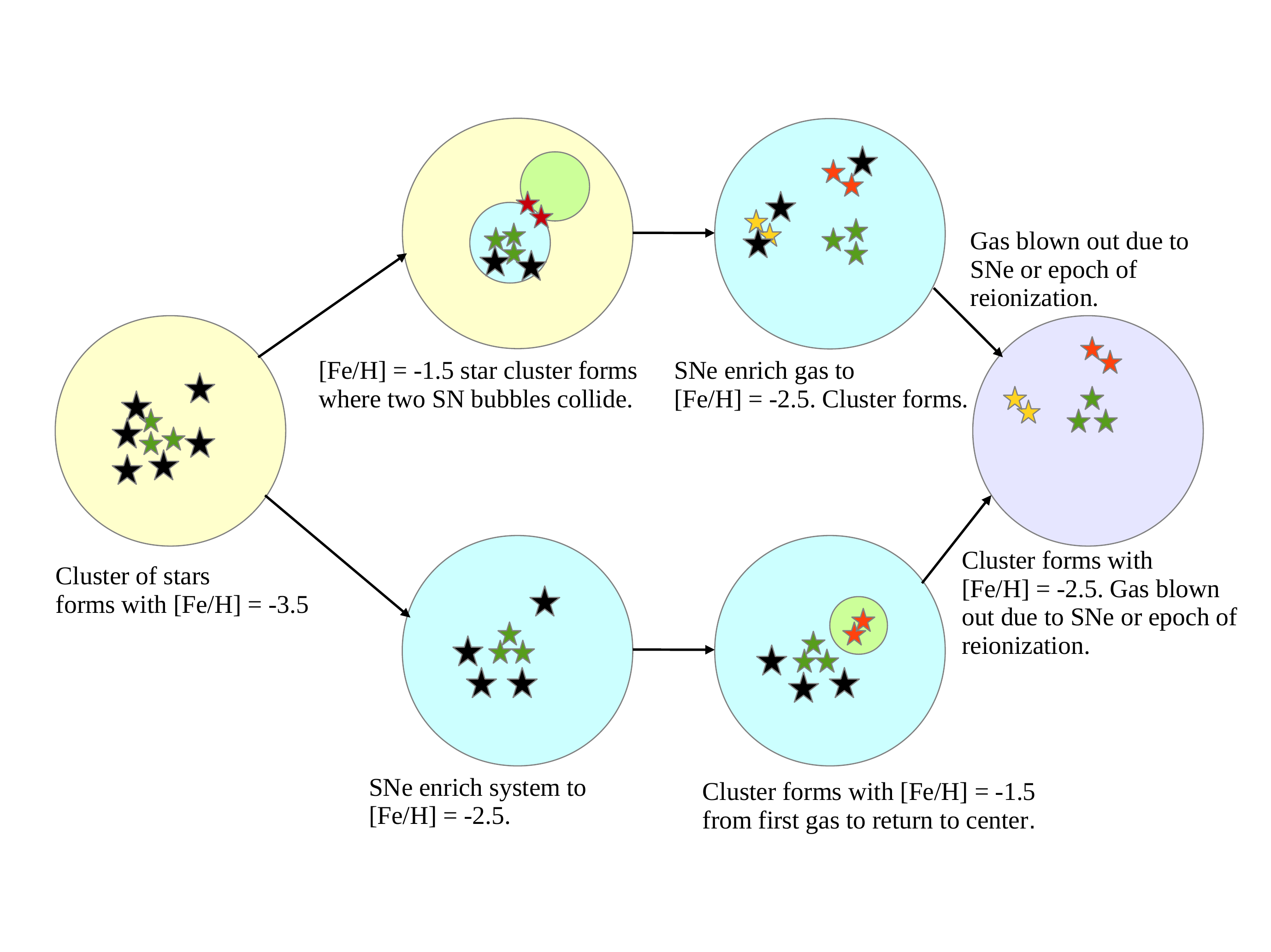}
      \caption{Two possible star formation histories for Segue~1. Both models start with a cluster or clusters of stars forming at [Fe/H]~$\approx-3.5$ 
(green stars) and finish with stars with the metallicities observed in Segue~1 today. The upper track has a cluster of stars forming at [Fe/H]~$\approx-1.5$ (red stars) at the collision of two supernova remnants. Subsequent supernovae enrich the gas to [Fe/H]~$\approx-2.5$, at which time another cluster of stars forms (yellow stars). Star formation is then 
terminated within 50~Myr of the first cluster forming. In the lower track, the gas is first enriched to [Fe/H]~$-2.5$ by $\sim10$ supernovae, with little star formation between them. There is then a gap between supernovae in which gas returns to the center. The first gas to return is more metal-rich and 
forms a cluster of stars with [Fe/H]~$\approx-1.5$ (red stars), with the cluster at [Fe/H]~$\approx-2.5$ (yellow stars) forming later when the gas is more 
well-mixed. Star formation is then terminated either by supernovae or a global event such as reionization.
}
     \label{f:cartoonplot}
\end{figure*}

Reaching an average [Fe/H]~$\approx-1.5$ with only Type II supernovae from the [Fe/H]~$\approx-3.6$ in the lowest metallicity three stars 
would require enrichment by $\approx70$ events per $10^5$~M$_\odot$ of gas. The stellar mass of 1500~M$_\odot$ in Segue~1 given a \citet{kroupa01} IMF therefore 
implies too few enrichment events. A top-heavy IMF \citep{geha13} could cause sufficient enrichment events, but it is likely that 70 supernovae per 
$10^5$~M$_\odot$ of gas within 50~Myr would blow out most of the metals and much of the gas \citep{maclow99,webster14}.  

Inhomogeneous mixing must therefore be at least part of the explanation for the highest metallicity stars. However, gas with metallicity [Fe/H]~$=-1.5$ 
rarely shows densities sufficient for star formation in gas accreted onto a dark matter halo \citep{greif10}, or after a single supernova event \citep{hawthorn15}. Any proposed star formation history must not only explain why there are no stars with metallicities between [Fe/H] $=-1.7$ and $-2.3$, or 
between [Fe/H]~$=-2.5$ and $-3.4$, but also how stars with such a high metallicity can form despite the short timescale of star 
formation in the system. 

We consider three possibilities to explain the MDF. In the first, the clusters of higher metallicity stars form at the interface of two colliding 
supernova remnants, where dense gas is swept up and partially mixed with the ejecta. The second scenario has the entire system enriched to [Fe/H]~$\approx-2.5$ 
by $\sim10$ SNe, with no star formation possible in between, because each supernova temporarily pushes out the dense gas. The [Fe/H]~$\approx-1.5$ stars form from the first dense gas that returns to the center in a pause between supernovae, with the [Fe/H]~$\approx-2.5$ stars forming shortly after. Finally, we consider the possibility of enrichment by exotic types of SNe such as pair-instability supernovae. 

\subsection{Simulations}
\label{s:simulations}

Our hydrodynamical simulations are described in detail in \citet{hawthorn15}. We used the Fyris Alpha hydrodynamical code \citep{sutherland10}, 
which solves ideal Euler hydrodynamical flows of gas with cooling and an equation of state appropriate to astrophysical gases in a fixed gravitational 
potential. A nested three-level grid was set up, with each level having 216$^3$ cells and being a factor of 3 smaller than its outer containing level. 
The largest level covers the halo out to beyond the virial radius of 630~pc, while the level containing the inner region has a resolution of 1.4~pc per cell. At $t=0$, a 25~M$_\odot$ star is assumed to form in a low mass ($M_{\rm{vir}}=10^6-10^7$~M$_\odot$) dark matter halo.

The effect on the gas, which was initially set up as a fractal interstellar medium in approximate equilibrium, was traced for 25~Myr with a 
resolution of 0.5~Myr. The models include the energy 
output of a 25~M$_\odot$ star both during the main sequence lifetime and from the supernova after 6~Myr. The ionisation phase from the precursor 
star was modelled using the MAPPINGS IV ionisation code \citep{allen08} for the thermal and ionisation structures, the ATLAS9 atmospheric grid \citep{castelli04}, and 
the \citet{meynet02} evolutionary tracks. The full details are in Section 2 of \citet{webster14}. The supernova was modelled by inserting a 
bubble of hot gas with an equivalent energy of $10^{51}$~erg. As the high-pressure bubble expands, its thermal energy is converted to kinetic energy and 
radiative losses. We assume the star explodes as a typical core-collapse supernova. Faint supernovae \citep[e.g.][]{cooke14} and other exotic supernovae 
will be discussed in Section 3.4.

The ionisation and supernova from a 25~M$_\odot$ star permanently switched off star formation for systems with $M\lesssim10^{6.5}$~M$_\odot$, but above this limit, dark matter halos retain sufficient baryons to continue forming stars. A star positioned away from the center couples less 
efficiently with the dense gas, with much of the energy escaping in the directions away from the center. This results in more gas retention and less enrichment. In a $10^7$~M$_\odot$ dark matter halo with a baryon fraction of 10\%, star formation is switched off for $\sim10$~Myr after the explosion 
of a high mass star near the center, but continues almost unaffected if the progenitor star is $\gtrsim80$~pc away from the center \citep{webster14}.

In this section we consider two variants of the above model as approximations to two possible star formation histories of Segue~1. The first involves the 
collision of a centered and off-centered explosion, while the second models a single supernova after the gas has been enriched to [Fe/H]~$=-2.5$. 

\subsection{Merging supernova remnants}

One possible scenario for the star formation history of Segue~1 involves the highest metallicity stars forming at the site where two supernova remnants 
collide. A single supernova expanding in an HII region can also induce star formation \citep{whalen08,nagakura09}, but here we investigate the colliding case, which allows for higher metallicities to be reached. This possibility was briefly considered in relation to first galaxies by \citet{greif07}, who noted that if a collision between SN remnants occurred early in their evolution 
in an overdense region, the density could become high enough for gravitational fragmentation and therefore star formation. 

As predicted by \citet{ostriker88}, a supernova remnant develops a dense shell of gas when its age is approximately equal to the cooling time of 
the shock-heated gas. Numerical simulations of the ISM of the Milky Way \citep{rosen95, korpi99,avillez00,joung06} have found that when these shells collide, dense and cold filamentary clouds are formed due to thermal instabilities and supersonic turbulence, an effect first noted for the case of two interacting supernova remnants by \citet{bodenheimer84}. In our simulations, given gas densities $\rho>1$~cm$^{-3}$, which are observed in the swept-up gas, the cooling time is shorter than our 500~kyr time resolution and star formation is therefore possible. If sufficient 
metal mixing occurs in the swept-up gas, the stars will form at a high metallicity. To investigate this scenario, we used the Fyris Alpha simulations 
discussed briefly in Section~\ref{s:simulations} and in detail in \citet{hawthorn15}.

The scenario is illustrated in the upper path of Figure~\ref{f:cartoonplot}. At $t = 0$, 
there is a dark matter halo containing gas enriched to [Fe/H]~$=-4$. This starting metallicity is uncertain and could be as high as [Fe/H]~$=-3.5$. In 
this gas, a cluster of stars forms. This cluster includes stars with masses $M\approx0.75$~M$_\odot$ observed as red giants today, 
denoted by the green stars, along with at least one star with $M>8~$M$_\odot$ (black stars in Figure~\ref{f:cartoonplot}). At a similar time, a cluster forms in lower density gas well away from the center with at least one massive star. The two clusters contain the 3 stars with 
metallicities of approximately [Fe/H]~$=-3.5$ observed as red giants today. 

\begin{figure}
     \centering
     \subfigure{
          \label{f:cartoonfeh}           
          \includegraphics[width=.45\textwidth]{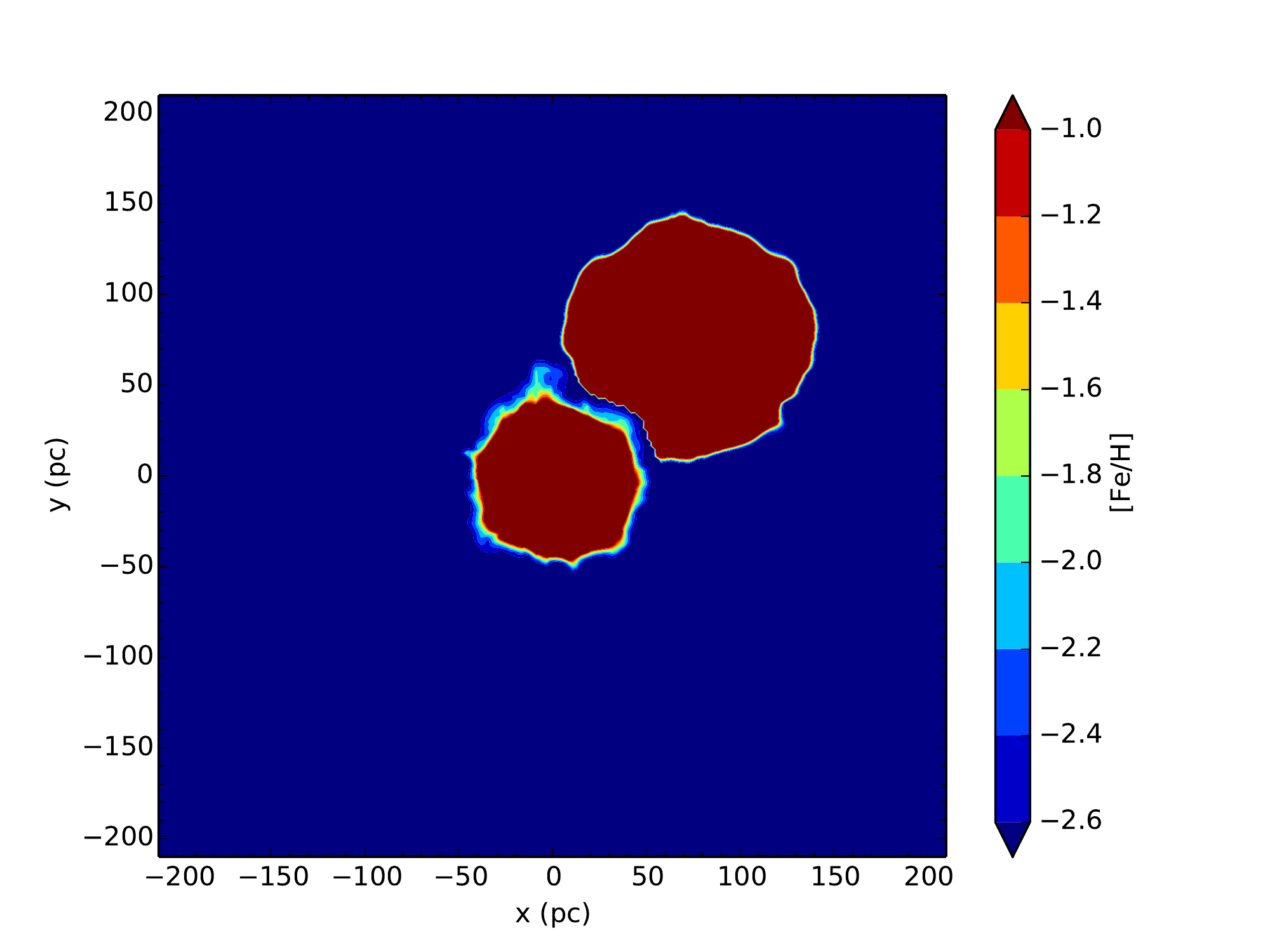}}
     \centering
     \subfigure{
          \label{f:cartoondens}
          \includegraphics[width=.45\textwidth]{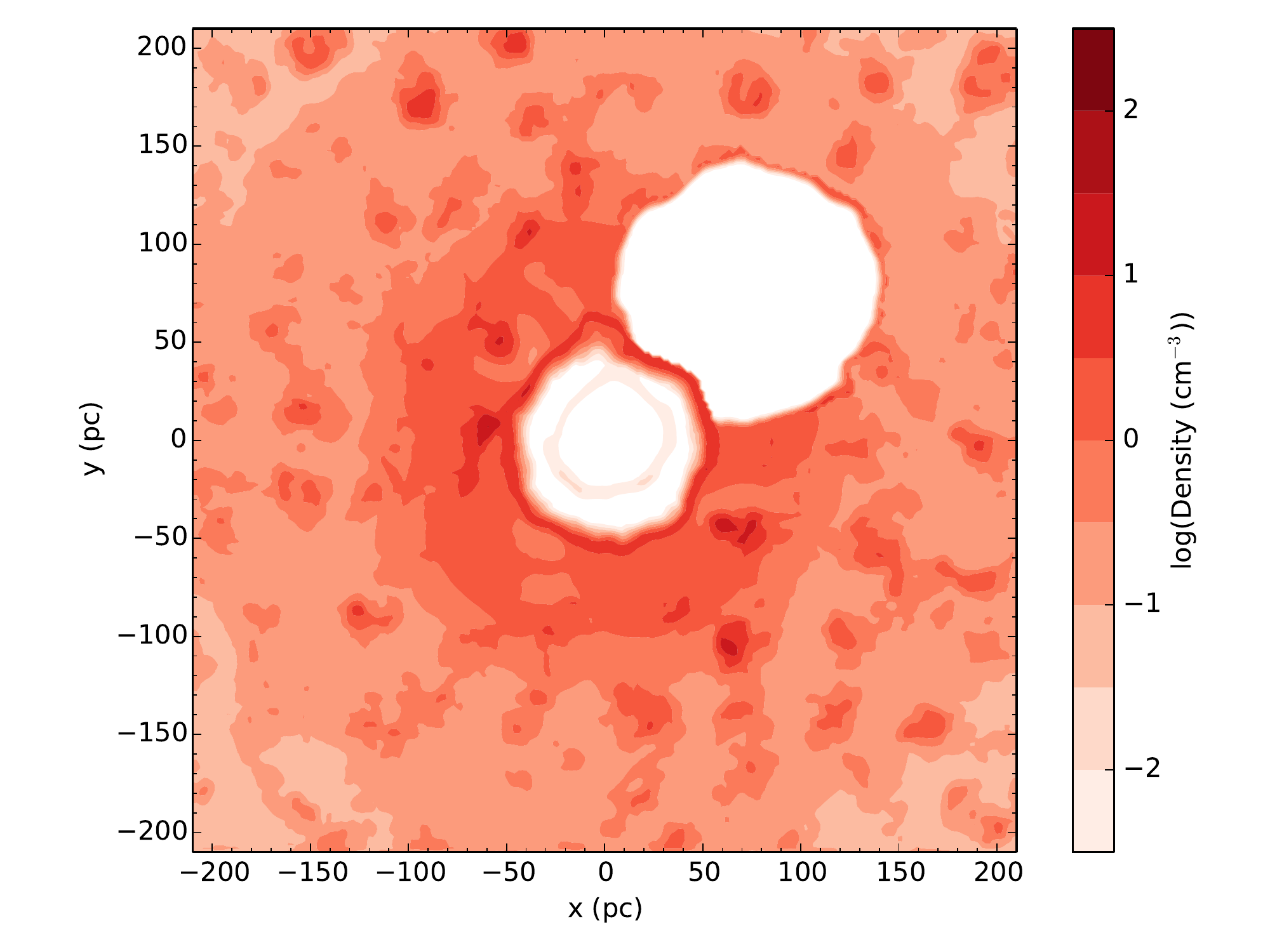}}
      \caption{A central slice of the metallicity and density of the gas just before the collision of supernova remnants from central and 
off-centered explosions. The two remnants have swept up dense neutral gas.}
     \label{f:mergsims}
\end{figure}

One supernova then occurs in each of the clusters described above. The inner supernova ejects 0.08~M$_\odot$ of Fe into a region with $2\times10^4$~M$_\odot$ of gas, enriching it to $\overline{\rm{[Fe/H]}}~=-2.5$. At the same time, the supernova further out, which occurs in less dense gas, has enriched a region with $2\times10^3$~M$_\odot$ of gas to $\overline{\rm{[Fe/H]}}~=-1.5$. These supernova remnants expand, sweep up gas and collide, inducing star formation in the swept-up gas. 
A cluster of stars forms with [Fe/H]~$=-1.5$ (red stars in Figure~\ref{f:cartoonplot}). The density and metallicity in the simulations at the time at which the two supernovae remnants collide is shown in Figure~\ref{f:mergsims}. Note that these simulations are unable to describe the complex physics 
of colliding supernova remnants, so here we are interested only in establishing whether the scenario is plausible.

Subsequent supernovae then enrich the gas to [Fe/H]~$=-2.5$ and a cluster forms at [Fe/H]~$\approx-2.5$ containing two low-mass stars (yellow in 
Figure~\ref{f:cartoonplot}) that will be red giants today. These stars must form less than 50~Myr after the first cluster of stars, because stars in 
Segue~1 do not show evidence of AGB or Type Ia enrichment. Finally, gas is blown out, terminating star formation. The gas can be blown out 
either by multiple closely-spaced supernovae, or as a result of the epoch of reionization. As noted by \citet{nichols11}, heated gas can be 
removed through tidal stripping by the host galaxy, or ram pressure stripping by a tenuous a hot halo.   


In the merging supernova remnants scenario, the reason that stars form at [Fe/H]~$\approx-1.5$ is that this is the 
metallicity of dense gas swept up and mixed with the supernova ejecta. In the future we will seek to perform hydrodynamical simulations of the effects of colliding supernova remnants, including 
whether the conditions for star formation exist, but this is beyond the 
scope of the present work. Here we simply note that at the boundary of the remnants, gas with [Fe/H] similar to the highest metallicity stars can 
build up, such that given the right conditions, stars will form. The stars at [Fe/H]~$\approx-2.4$ may be formed slightly later at another site along the boundary between the two remnants, or by the first dense gas that 
returns to the center of the system after $\sim10~$Myr, which is more metal-rich than the mean metallicity, as we will see in Section~\ref{s:twoburst}. 


We therefore suggest the following explanation for the MDF in Segue~1. The first stars formed with [Fe/H]~$\approx-3.6$, near the mean 
metallicity of the system when it formed. The first two SNe occurred at approximately 
the same time, separated by at least $\sim75$~pc. Dense gas was swept up and partially mixed with the supernova ejecta. A shock formed and the gas 
cooled, resulting in a cluster of stars forming with [Fe/H] much higher than the mean metallicity of the system. All the star formation in the 
system could be complete within 10-15~Myr, before supernova feedback blew all the gas out. A top-heavy initial mass function as suggested by \citet{geha13} 
is expected to produce 250-400 supernovae in Segue~1 \citep{frebel14}, which could easily blow the gas out of an $M_{\rm{vir}}\sim10^7~$M$_\odot$ 
dark matter halo. 


This scenario is particularly interesting in light of the \citet{ritter15} simulations discussed in Section 2.4. They found 
enhanced carbon relative to iron for gas which falls into the center. The iron-rich ejecta has more entropy and reaches higher radii than the 
carbon-rich ejecta, resulting in the stars which form at the center being carbon enhanced. Here we instead have stars forming at higher radii where two 
supernova remnants merge. If the iron has overtaken the carbon by the time of the collision, the stars which form will have high [Fe/H] and low [C/Fe]. 
This is observed in the star at [Fe/H]~$=-1.5$ with [C/Fe]~$\sim0$. It should be noted that \citet{ritter15} model 
lower-mass ($10^6$~M$_\odot$) systems over longer time periods, but it is suggestive of a possibility for inhomogeneous enrichment.     

\subsection{Two bursts}
\label{s:twoburst}

Another possible star formation history for Segue~1 is shown in the lower track of Figure~\ref{f:cartoonplot} and can be summarised as follows:
\begin{enumerate}
\item  A cluster of second-generation stars formed at $t=0$ with [Fe/H]~$\approx-3.5$ (shown in green in Figure~\ref{f:cartoonplot}).
\item  Within the 30-50~Myr allowed for star formation, $\approx10$~SNe resulted from the star formation of this first cluster, enriching the medium to [Fe/H]~$\approx-2.5$. This supernova rate is too high for the gas to recover between events, so little or no star formation occurs during this time.
\item Due to the stochasticity of supernova events at low star formation 
rates, there was a sufficiently long gap $T\sim10$~Myr between supernovae, allowing the gas to return to the center. The first gas to reach the center can have metallicities as high as [Fe/H]~$\approx-1.5$ and reach sufficient densities that stars may form (red stars in 
Figure~\ref{f:cartoonplot}. Gas in different parts of the system remained at [Fe/H]~$\approx-2.5$ and formed another cluster (yellow stars in 
Figure~\ref{f:cartoonplot}).
\item  As a result of the star formation from these clusters, or due to the epoch of 
reionization or other environmental effects, the star forming gas was evaporated or blown out of the system, resulting in the termination of star 
formation.
\end{enumerate}
  
We investigated this possibility using the Fyris Alpha simulations described in Section 3.1 and in \citet{hawthorn15}. However, instead of a 
uniform starting metallicity [Fe/H]~$=-4$, we assumed a starting metallicity of [Fe/H]~$=-2.5$, such that the simulation models only the 
final supernova in step 2 of the scenario discussed above. In the initial state of the simulation, we assume that all the gas has this starting metallicity. In reality, it is likely that the gas will not be this well-mixed, and that regions of higher metallicity will exist. 
These pre-existing inhomogeneities may assist in reaching metallicities higher than 
we find.

We assume that a supernova occurs in a medium already somewhat disrupted by the previous supernovae. 
While \citet{hawthorn15} models a 25~M$_\odot$ star, here the star will be lower 
mass, because $\sim10$ supernovae from the cluster have already occurred, suggesting a time scale of $20-30$~Myr. This implies a mass of $<15$~M$_\odot$. 
However, the effect of the previous supernovae will be such that the surrounding medium is likely to be largely evacuated. As an approximation, the 
supernova is therefore assumed to occur in the same medium as that for a 25~M$_\odot$ star.

\begin{figure}
     \centering
     \subfigure{
          \label{f:simsfeh}           
          \includegraphics[width=.45\textwidth]{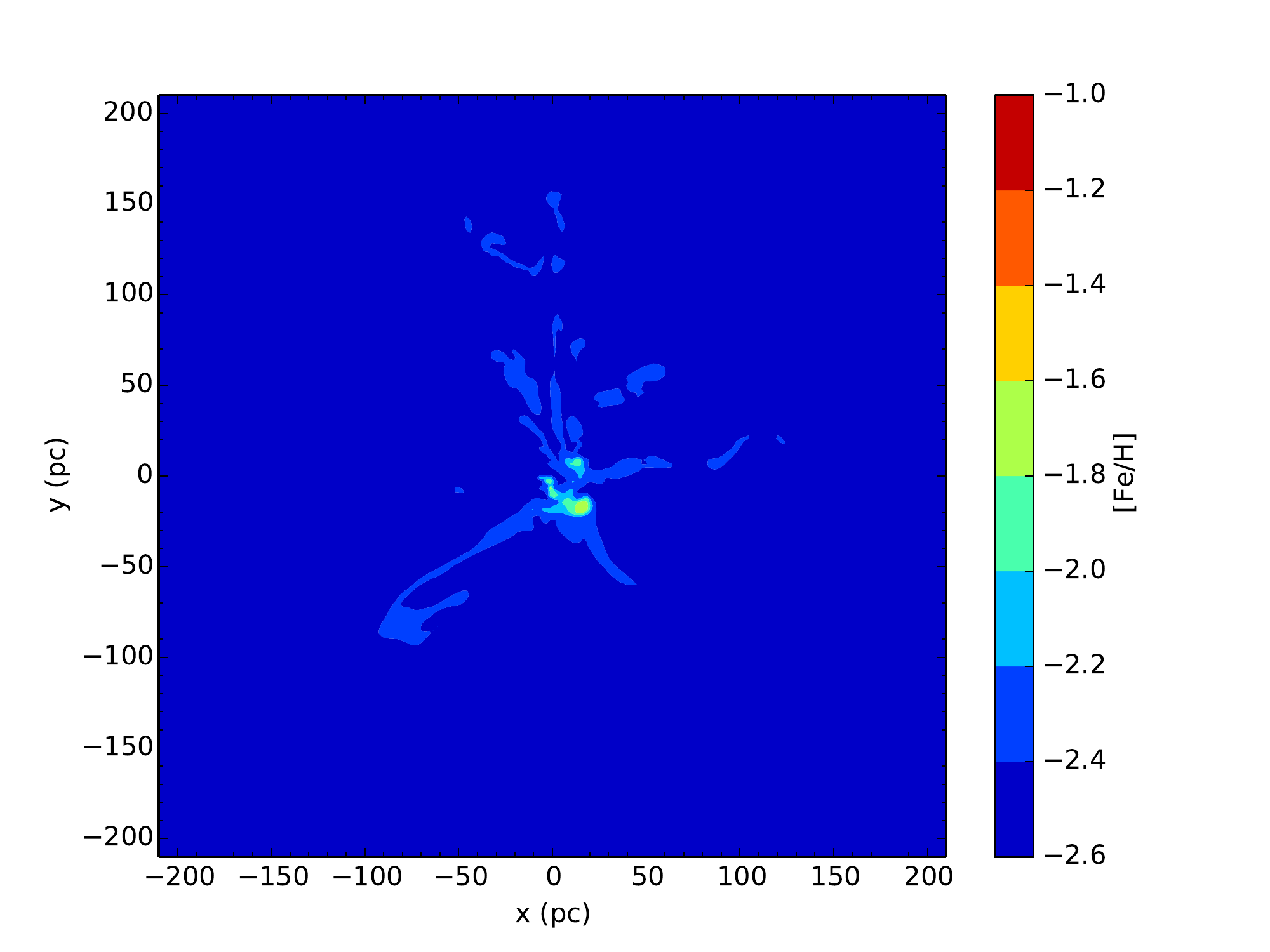}}
     \centering
     \subfigure{
          \label{f:simsdens}
          \includegraphics[width=.45\textwidth]{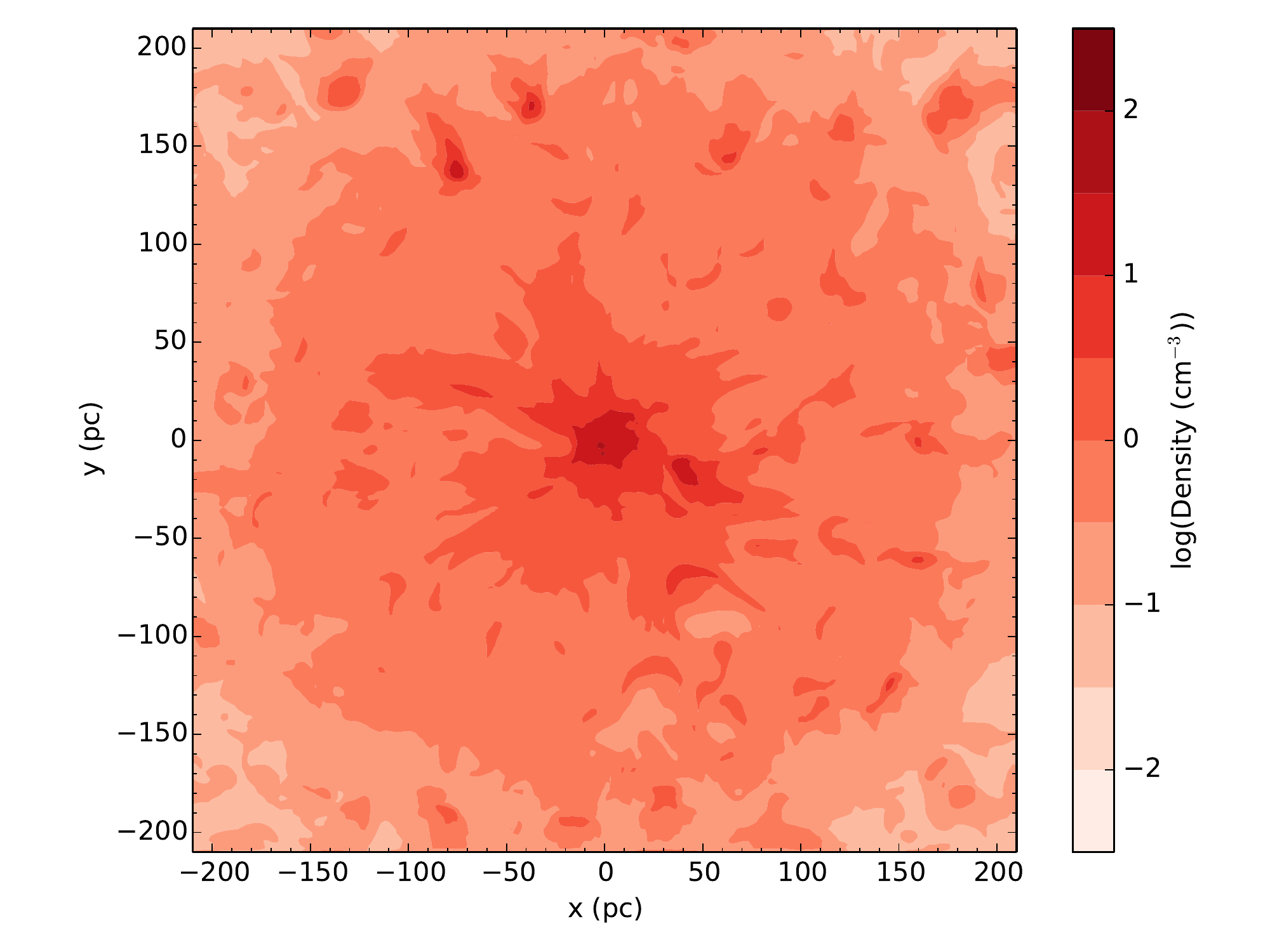}}
      \caption{A central slice of the metallicity and density of the gas 11~Myr after a supernova, the time at which the first dense gas is 
returning to the center.}
     \label{f:sims}
\end{figure}

The effect of the supernova is shown in Figure~\ref{f:sims}. The gas that returns to the center after $\sim10$~Myr is enriched to higher [Fe/H] then 
the overall average. While the metallicity of the gas in the central 20~pc has mean [Fe/H]~$\approx-2.0$, there are pockets of dense gas with metallicities 
[Fe/H]~$\gtrsim-1.6$. For example, 11~Myr after the supernova, there is a clump of gas with a mass of 200~M$_\odot$ and [Fe/H]~$=-1.6$ within an approximately spherical region with a radius of 7~pc. 

The third cluster observed today may form at a similar time in one of the lower metallicity clumps. The combined effect of the supernovae from these 
two newly-formed clusters, or the onset of the epoch of reionisation, may then eject all the gas in the system such that the three clusters described above contain all the stars that ever formed in Segue~1.

This scenario provides a natural explanation of the observed MDF. The first cluster at [Fe/H]~$\approx-3.5$ is responsible for the three lowest metallicity 
stars observed as red giants today. The supernovae resulting from this cluster enrich the ISM to [Fe/H]~$\approx-2.5$, while blowing the gas out of the 
center such that stars with metallicities in the range $-3.5<$~[Fe/H]~$<-2.5$ are prevented from forming. The other two groups of stars form during a gap 
between supernovae, before star formation is permanently switched off.

The abundances of elements relative to iron should then be close to the average Type II supernova yields, although there may be a bias towards 
the higher mass stars, which explode first. The first 10 supernovae may not be the entire supernova output of the first cluster, in which case the 
lower mass supernova progenitors would not yet have exploded by the time the higher metallicity clusters in Segue~1 form. This is indeed seen, with 
[$\alpha$/Fe]~$\sim0.5$.

Under this scenario, part of the explanation for Segue~1 is that it has just the right mass and supernova rate that star formation can be temporarily 
switched off by supernova feed back without the gas being blown away completely. Such a scenario may be rare, which is why we have seen it only once 
in the UFDs discovered so far.

\subsection{Exotic supernovae}

A possible explanation for high [Fe/H] in a first galaxy is a pair-instability supernova (PISN). A single PISN in a precursor minihalo could enrich a halo with $10^5$~M$_\odot$ to an average [Fe/H]~$\approx-2.5$. \citep{heger03} There may then be sufficient natural scatter to explain the 
existence of stars at [Fe/H]~$\approx-3.5$ and $-1.5$. The suppressed [Ba/Fe] and [Sr/Fe] in Segue~1 also supports this explanation. However, measurements 
of $\alpha$ abundances, as well as [Co/Fe] \citep{norris10b,frebel14}, do not match the distinctive signatures expected for a PISN \citep{heger02}. The suggestion that different elements mix differently into the ISM \citep{sluder15,ritter15} may be a partial solution to this, however this effect does not explain why all the stars in Segue~1 show similar abundance patterns for all $\alpha$ and Fe-peak elements.

Other alternatives to regular core-collapse supernovae do not match the abundance patterns of Segue~1. Pulsational pair-instability 
supernovae eject mostly lighter elements, with the iron lost as the star ends its life as a black hole \citep{heger03}, so are not a solution to the observed high [Fe/H]. Hypernovae can eject more iron into the ISM than regular supernovae, but only by a factor of $2-3$ \citep{nomoto06}. They are 
more energetic than regular supernova by an order of magnitude, so will blow out a large proportion of gas and metals from a low mass system. Faint 
supernovae, which produce small amounts of iron compared to carbon have been suggested by \citet{cooke14} as an explanation for the formation of 
CEMP stars in minihalos, as their lower energy means they are less efficient at blowing out gas. However, they eject only a small amount of iron and would therefore struggle to 
explain the higher-metallicity stars in Segue~1.

It is therefore unlikely that exotic types of supernovae are responsible for the observed abundances of Segue~1. However, a pair-instability 
supernova could produce a similar [Fe/H] distribution, so should be considered if other systems showing similar MDFs to Segue~1 are discovered.

\section{Discussion}
\label{s:disc}

\subsection{The nature and enrichment history of Segue~1}

Segue~1 provides a unique opportunity to investigate the early metal build-up in galaxies because it experienced so few chemical enrichment events. 
In all other known galaxies, enrichment from later generations 
complicates the picture. UFDs were likely some of the early building blocks of the Milky Way halo, so by studying galaxies like Segue~1, we can help  determine the chemical history of the Milky Way.

Our results point to a new way of looking at metal build-up. When there are few previous enrichment events, abundances at a given location can 
be dominated by the effect of the last supernova. [Fe/H] and [C/H] do not increase uniformly everywhere in a galaxy, but are instead highly 
inhomogeneous, with some places reaching high metallicity before further mixing takes them back to close to the average metallicity for the galaxy. This ``two steps forward, one step back" enrichment means that some relatively metal-rich stars form before lower-metallicity stars. 

Therefore, while after many supernovae there is a gradual increase in average metallicity over time, this is not necessarily the case locally and over 
shorter periods of time. Hence, the oldest stars are not necessarily the most metal-poor stars, and more importantly, ancient metal-rich stars are expected 
to exist. None of the scenarios that we consider plausible for Segue~1 has the lowest metallicity stars enriching the gas which forms 
stars with moderate metallicity, which then enriches the gas that forms the highest metallicity stars. We find that the most likely scenarios have either the lowest 
metallicity [Fe/H]~$\approx-3.6$ stars enrich the gas that forms both the higher metallicity clusters, which share the same enrichment history, or that 
all stars in Segue~1 were formed from gas with the same enrichment history. Furthermore, given that they require metallicities much higher than the 
likely average of the gas in Segue~1, the [Fe/H]~$\approx-1.5$ stars are more likely to have formed before the [Fe/H]~$\approx-2.4$ stars. Gaia will provide accurate ages for very metal-poor halo stars and can test for such out of order enrichment. If our view of Segue~1 and the 
chemical enrichment of early systems is correct, we predict that some of the oldest stars may have higher metallicities [Fe/H]~$>-2.0$. 

In systems with few enrichment events, short-term variations in environment can result in a large variance in [Fe/H]. The first stars in the 
system can therefore be outliers in chemical abundance space. However, if star formation is 
clustered, groups of stars form with very similar abundances. The early outliers in abundance space resulting from the varied environment come in groups of 
stars rather than individually. This means that rather than each individual system having a small number of outlier 
stars, some systems may have a large number (if they contain a cluster or clusters which is an outlier in abundance space) and others might have 
none (if the early clusters are not outliers). In this way, 
we explain how 5 of the 7 stars in Segue~1 can have $\Delta$[Fe/H]~$>0.8$~dex from the mean, whereas most other ultra-faint dwarfs (UFDs) have much narrower MDFs. The MDFs seen in the majority of UFDs can result either from the system having a large number of enrichment events, 
filling in the MDF and hiding distinct clusters, or by only having a few clusters with similar [Fe/H]. The latter class of systems do not form
outlier stars at any time. 

For most UFDs, the overall picture is one of early environment-dependent enrichment, followed by increasing homogeneity after sufficient enrichment events have occurred. 
In this way, we can account for both overall homogeneity in the abundances of most stars \citep{cayrel04}, along with the unusual abundances of others \citep{frebel15,hansen15}. The unusual stars, which are not necessarily the most metal-poor, are those which formed first.

Our work is in line with that of \citet{ji15}, who suggested that CEMP stars are second-generation stars enriched only by massive Pop III stars. Subsequent 
generations quickly wash out this carbon enhancement. This fits with Segue~1, where the stars with [Fe/H]~$\approx-3.5$ are CEMP stars, while the higher 
metallicity stars, which in our models are formed from gas enriched by these second-generation stars, all have [C/Fe]~$\approx0$. Furthermore, it is another 
example of outlier stars being the first to form. The yields of Population III stars appear to be determined best by studying stars in dwarf galaxies and the 
Milky Way halo which have unusual abundances.   

The discovery of $\sim20$ new UFDs in 2015 is likely to help answer the question of whether Segue~1 is a rare system, or its chemical abundances are 
common for the faintest galaxies. Its MDF is unusual compared to all other known UFDs, but nearly all of them are more luminous systems with longer 
star formation histories.

\subsection{Reticulum 2}

The UFD Reticulum 2 is similar to Segue~1 in its distance, mean metallicity and luminosity. The system was discovered only recently, so at the time of writing only iron \citep{simon15,walker15,koposov15b} and medium-resolution ($R\sim18000$) $\alpha$ abundances \citep{koposov15b} have been determined. As shown in Figure~\ref{f:theoscen}, its MDF appears similar to the more luminous UFDs, contrary to Segue~1. However, this does not rule out a similar star formation history to Segue~1. 

In our scenarios above, the [Fe/H]~$=-2.4$ stars in Segue~1 formed from gas enriched by the first cluster of stars at [Fe/H]~$=-3.5$. This produced a gap 
in the MDF, but equally plausible scenarios may not form such a gap. Because of the logarithmic nature of [Fe/H], a first cluster at [Fe/H]~$=-3.0$ would 
require $>75\%$ of the number of supernovae to reach [Fe/H]~$=-2.5$ as a first cluster at [Fe/H]~$=-3.5$. It would be much more difficult to 
distinguish between two clusters with means separated by 0.5~dex than two clusters separated by 1~dex. 

The scenarios in Section~\ref{s:scen}, may not always lead to the formation of stars with [Fe/H]~$\approx-1.5$ will form. 
Star formation in merging supernova 
remnants may not be common, or the gas which returns to the center first may not form stars (or, as in \citet{ritter15}, this gas may be carbon-rich and iron-poor). It is therefore possible that we will find many more examples of Ret 2-like MDFs than Segue~1, even if both are first galaxies, and similar processes occur in both systems. The new galaxies discovered by DES provide an excellent opportunity to test this. 

There are multiple possible explanations for the origin of Ret 2. The first is that it has a similar history to Segue~1, with clusters of stars forming 
at [Fe/H]~$\approx-3.2$ and [Fe/H]~$\approx-2.5$, but no higher metallicity cluster (or a small high-metallicity cluster from which we have not 
yet observed stars). Alternatively, the MDF of Ret 2 is consistent with a gradual build-up of metallicities as in the ``Type II" model described in 
Section 2.1, or at the other extreme, a single burst with $\overline{\rm{[Fe/H]}}\approx-2.7$ and $\sigma\approx0.3$. In Segue~1 we are fortunate that the clusters are well-separated, allowing us to distinguish between these models.

Ret 2 is the closest of these new systems, at 30~kpc \citep{koposov15a} and is therefore the best candidate for performing high-resolution abundance measurements. High-resolution spectroscopy is required to determine whether Ret 2 passes the tests for a ``first galaxy" as outlined in \citet{frebel12}. A first galaxy should not show stars, even at [Fe/H]~$>-2$, with [$\alpha$/Fe] systematically lower than the halo value of 0.35. Furthermore 
there should be no signature of s-process or carbon enrichment from AGB stars. Medium-resolution observations of Ret 2 \citep{koposov15b} provide a tentative suggestion that [$\alpha$/Fe] does not decline with increased 
[Fe/H], with all 17 observed stars showing [$\alpha$/Fe]~$>0.2$, close to the mean [$\alpha$/Fe]~$=0.4$. High-resolution observations are required to test this more rigorously, as well as to determine the presence or absence of neutron-capture elements. 

Segue~1 is highly unusual compared to all other observed galaxies. However, first galaxies may be 
difficult to observe because they are likely to be intrinsically less luminous due to their shorter period of star formation, so this could be a 
result of an observational bias. If Ret 2 proves to be similar to Segue~1, there will be two known first galaxies, helping determine which 
features are typical of first galaxies and which are unique to individual systems.

\section{Summary}
\label{s:summary}

We have used the metallicity distribution function of Segue~1 to reconstruct its the star formation history and to learn about the nature of 
first and early galaxies. In Section~\ref{s:theomdfs}, we showed that Segue~1 is best explained by clustered star formation. 
A gradual increase in [Fe/H] with time can 
not reproduce its MDF. There are three distinct groups in [Fe/H], with gaps of $1.2$ and $0.55$~dex. All other UFDs 
observed to date have 90\% of their stars within 0.5~dex of the mean [Fe/H] for the system and few show stars with metallicities as low as 
[Fe/H]~$\approx-3.6$ or as high as [Fe/H]~$\approx-1.5$ as in Segue~1. We conclude that the unusual MDF is a sign of earliest star formation, in line with Segue~1 being a ``first galaxy", meaning that it experienced only a single burst of star 
formation lasting $\lesssim50$~Myr. There are a number of features of Segue~1 that are naturally explained by the first galaxy hypothesis:

\begin{enumerate}
\item The gaps in the MDF are caused by the stochasticity within a single generation of star formation. Segue~1 is the only known galaxy which appears 
not to have experienced self-enrichment and therefore multiple generations of star formation. Multiple generations of star formation are likely to wash 
out stochastic effects, resulting in the MDF being filled in near the average metallicity.
\item The highest metallicity stars remain difficult to explain, but may be related to the large variation in metallicity caused by even a single 
supernova in a low-mass system. First galaxies are more likely to have low masses, because they are more susceptible to effects such as supernova 
feedback and the epoch of reionization, which can quench their star formation.
\item Segue~1 has the lowest intrinsic luminosity of any known galaxy. A Kroupa IMF implies that it has only formed only $\approx1500$~M$_\odot$ of stars over its lifetime. This low amount of total star formation is likely related to the short duration of star formation.  
\end{enumerate}

In Section~\ref{s:scen}, we used hydrodynamical simulations to test two possible explanations for the MDF of Segue~1. One possibility is that the 
high-metallicity stars formed from swept-up gas at the interface between two colliding supernova remnants. 
In this scenario, clusters of stars with a difference in [Fe/H] of 2~dex can form within 10~Myr of each other. Furthermore, as shown by 
\citet{ritter15} and discussed in Section 2.4, there can be large variance in [C/Fe] at early times after a supernova explosion in a low-mass system. 
In particular, the iron-rich ejecta has higher entropy and can overtake the carbon-rich ejecta. If this occurs by the time of the collision, it provides 
an explanation for [C/Fe] being suppressed by $1-2$~dex for the stars at [Fe/H]~$\approx-1.5$ compared to those at [Fe/H]~$\approx-3.5$.

Alternatively, our hydrodynamic simulations show that in an $M_{\rm{vir}}=10^7$~M$_\odot$ dark matter halo with $10^6$~M$_\odot$ of gas 
pre-enriched to [Fe/H]~$=-2.5$, the first gas that returns to the center of the system after a supernova explosion will be enriched to [Fe/H]~$\approx-1.5$ 
and be sufficiently dense that star formation is plausible. We also briefly explored the possibility of more exotic types of supernovae such as PISNs to reproduce the MDF of Segue~1, but these could not explain the abundances of other elements.

At the time of writing, 17 new UFD candidates have been detected in DES \citep{koposov15a,des15}. One of these, Reticulum 2, which has been 
confirmed as a galaxy, is at a similar distance as Segue~1 and has a similar mean metallicity. Medium-resolution observations of [$\alpha$/Fe] suggest 
it as a tentative candidate for a first galaxy. High-resolution abundance measurements are required to determine whether this is indeed the case. Segue~1 
is the only known first galaxy and it is therefore currently impossible to say whether its features are typical or unusual. The newly-discovered systems 
provide an exciting opportunity to test this. In the long-term, LSST will be capable of detecting much fainter objects, while G-CLEF on the GMT will enable 
high-resolution abundance measurements of fainter stars.

\acknowledgements

DW is supported by an ARC Laureate Fellowship awarded to JBH.  We acknowledge the Dick Hunstead Fund for Astrophysics for funding
AF's visit to the University of Sydney to work on this paper. AF acknowledges support from the Silverman (1968) Family Career Development Professorship. AF is supported by NSF grant AST-1255160. This research was also supported in part by the National Science Foundation under Grant No. NSF PHY11-25915, as 
the collaboration was formed at the Kavli Institute at UC Santa Barbara during the Galactic Archaeology and Precision Stellar Astrophysics program. We thank the anonymous referee for comments which helped improve this work.

\bibliographystyle{apj}
\bibliography{refs}

\end{document}